\documentclass{article}
\usepackage[margin=0.5in]{geometry}
\usepackage{graphicx}
\usepackage{amsmath}
\usepackage[obeyspaces]{url}
\PassOptionsToPackage{obeyspaces}{url}
\usepackage{hyperref}
\usepackage{authblk,color}
\usepackage{algorithm,amssymb}
\usepackage[english]{babel}
\usepackage{multirow}
\usepackage{blkarray}
\usepackage{caption} 
\usepackage{gensymb}
\usepackage{adjustbox}
\usepackage{listings}

\usepackage{setspace}

\begin{document}

\title{HuGaDB: Human Gait Database for Activity Recognition from Wearable Inertial Sensor Networks}
\author[1]{Roman Chereshnev}
\author[1]{Attila Kert\'esz-Farkas\footnote{Corresponding author at: Kochnovskiy Proezd, 3, Moscow, 125319, Russian Federation \newline Tel.: +7 499 152 0741 \newline 
		{\it E-mail addresses}: rchereshnev@hse.ru (R. Chereshnev), akerteszfarkas@hse.ru (A. Kert\'esz-Farkas)}}

\affil[1]{School of Data Analysis and Artificial Intelligence, Faculty of Computer Science, National Research University Higher School of Economics (HSE)}





\maketitle

\begin{abstract}
	This paper presents a human gait data collection for analysis and activity recognition consisting of continues recordings of combined activities, such as walking, running, taking stairs up and down, sitting down, and so on; and the data recorded are segmented and annotated.  Data were collected from a body sensor network consisting of six wearable inertial sensors (accelerometer and gyroscope) located on the right and left thighs, shins, and feet. Additionally, two electromyography sensors were used on the quadriceps (front thigh) to measure muscle activity.  This database can be used not only for activity recognition but also for studying how activities are performed and how the parts of the legs move relative to each other. Therefore, the data can be used (a) to perform health-care-related studies, such as in walking rehabilitation or Parkinson's disease recognition, (b) in virtual reality and gaming for simulating humanoid motion, or (c) for humanoid robotics to model humanoid walking. This dataset is the first of its kind which provides data about human gait in great detail. The database is available free of charge \url{https://github.com/romanchereshnev/HuGaDB}.	
\end{abstract}

\section{Introduction}
\label{intro}
The increasing availability of wearable body sensors leads to novel scientific studies and industrial applications  \cite{aggarwal2013managing}. The main large areas include gesture recognition, human activity recognition, and human gait analysis. Several databases have been released for benchmarking; however, due to a wide variety of sensor types and the complexity of activities, these databases are rather distinct. Now, we will review these areas and the corresponding databases in a taxonomic manner. 

{\bf Gesture recognition} (GR) mainly focuses on recognizing hand-drawn gestures in the air. Patterns to be recognized may include numbers, circles, boxes, or Latin alphabet letters. Prediction is usually made on data obtained from smartphone sensors or some special gloves equipped with kinematic sensors, such as 3-axis accelerometers, 3-axis gyroscopes, and occasionally electromyography (EMG) sensors, to measure the electrical potential on the human skin during muscular activities \cite{amma2014airwriting}. A database for gesture recognition is available in \cite{georgi2015recognizing}.

{\bf Human activity recognition} (HAR), on the other hand, aims at recognizing daily lifestyle activities. For instance, an interesting research topic is recognizing activities in or around the kitchen, such as cooking; loading the dishwasher or washing machine; preparing brownies or salads; scrambling eggs; light cleaning; opening or closing drawers, the fridge, or doors; and so on. Often these activities can be interrupted by, for example, answering phones.   Databases on this topic include the MIT Place dataset  \cite{tapia2006design,intille2005living}, Darmstadt Daily Routine dataset \cite{huynh2008discovery}, Ambient Kitchen \cite{pham2009slice}, CMU Multi-Modal Activity Database (CMU-MMAC) \cite{de2008guide}, and Opportunity dataset \cite{Chavarriaga2013,sagha2011benchmarking}. In this topic, on-body inertial sensors are usually worn on the wrist, back, or ankle, however, additional sensors are used, such as temperature sensor, proximity sensor, water consumption sensor, heart rate and so on. For instance, CMU-MMAC includes videos, audios, RFID tags, motion capture system based on on-body markers, and physiological sensors such as galvanic skin response (GSR) and skin temperature, which are all located on both forearms and upper arms, left and right calves and thighs, abdomen, and wrists.

Other types of HAR usually focus on walking-related activities, such as walking, jogging, turning left or right, jumping, laying down, going up or down the stairs, and so on. Data on this topic can be found in the WARD dataset \cite{yang2009ward}, PAMAP2 dataset \cite{reiss2012introducing,reiss2012creating}, HASC challenge \cite{kawaguchi2011hasc,kawaguchi2012hasc2012corpus,kawaguchi2011hasc2011corpus}, USC-HAD \cite{zhang2012usc,zhang2013human}, and MAREA \cite{khandelwal2017evaluation}. For data collection, on-body sensors are often placed on the participant's wrist, waist, ankles, and back.

In some databases, exceptional efforts are taken to provide a reliable benchmark. The body sensor network conference (BSNC) (http://bsncontest.org) \cite{giuberti2011simple}, for instance, has carried out a contest where organizers provided three different datasets from different research groups. Databases differ in sensor types used and activities recorded. Another team, called the Evaluating Ambient Assisted Living Systems through Competitive Benchmarking -- Activity Recognition (EvAAL-AR), provides a service to evaluate HAR systems live on the same activity scenarios performed by an actor \cite{gjoreski2015competitive}.   In this contest, each team brings its own activity recognition system, and the evaluation criteria attempt to capture the practical usability: recognition accuracy, user acceptance, recognition delay, installation complexity, and interoperability with ambient-assisted living systems.


{\bf Gait analysis} focuses not only on the recognition of activities observed but also on how activities are performed. This can be useful in health-care systems for monitoring patients recovering after surgery or fall detection or in diagnosing the state of, for example, Parkinson's disease \cite{sant2011new,sant2012symbolic}. For instance, the Daphnet Gait dataset (DG) \cite{bachlin2009potentials} consists of recordings of 10 participants affected with Parkinson’s disease instructed to carry out activities that are likely to be difficult to perform, such as walking. The objective is to detect these incidents from accelerometer data recorded from above the ankle, above the knee, and on the trunk. On the other hand, Bovi et al. provide a gait dataset collected from 40 healthy people with various ages as a reference dataset \cite{bovi2011multiple}. In the aforementioned BSNC, the third database (ID:IC) contains gait data before knee surgery and 1, 3, 6, 12, and 24 weeks (respectively) after it.

\section{Motivation and Design Goals}
\label{sec:motivation}
The main purpose of this dataset is to provide detailed gait data to study how the parts of the legs move individually and relative to each other during activities such as walking, running, standing up, and so on.  A summary of the activities can be found in Table \ref{tb:activity-data-dist}. This dataset contains continuous recordings of combinations of activities, and the data are segmented and annotated with the label of the activity currently performed. Thus, this dataset is also suitable for analyzing human gait and activities between transitions. 

Mainly inertial sensors were used for data acquisition. We decided to use inertial sensors because they are inexpensive, simple to use anywhere such as indoor and outdoor area, and widely available compared with other systems. For instance, compared with video-based motion capture systems, they require expensive video cameras and special full bodysuit with special markers on it. In addition, they are restricted to being used in the installed test area and they are sensitive to lightning and suffer from lost markers phenomenon. 

In total, six inertial sensors were placed on the right and left thighs, shins and feet; and data were collected from 18 healthy participants, providing total 10 hours of recording. This allows one to investigate how the parts of the legs move individually and relative to each other within and in-between activities. Our dataset could be used as control data, for instance, in health-care-related studies, such as walking rehabilitation or Parkinson's disease recognition. In virtual reality or gaming, our dataset can be used to model a virtual human movements by reproducing the leg movements from the accelerometer data by simply taking the integrals. In fact, it is not limited to virtual environment and could be used to train to walk and move humanoid robots to make them more humanlike and cope with the uncanny valley.

This dataset is unique in the sense that it is the first to provide human gait data in great detail mainly from inertial sensors and contains segmented annotations for studying the transition between different activities. 

\begin{table}[h]
	\caption{Characteristics of HuGaDB}
	\label{tb:activity-data-dist}
	\centering
	\begin{adjustbox}{max width=\textwidth}
		\begin{tabular}{llllll}
			\hline
			ID& Activity &Time sec (min) &Percent &Samples &Description\\
			\hline
			1&Walking &11544 (192) &32.15 &679073&Walking and turning at various speeds on a flat surface\\
			2&Running &1218 (20) &3.39 &71653&Running at various paces\\
			3&Going up &2237 (37) &6.23 &131604&Taking stairs up at various speeds\\
			4&Going down &1982 (33) &5.52 &116637&Taking the stairs down at various speeds and steps\\
			5&Sitting &4111 (68) &11.45 &241849& Sitting on a chair; sitting on the floor not included\\
			6&Sitting down &409 (6) &1.14 &24112& Sitting on a chair; sitting down on the floor not included\\
			7&Standing up &380 (6) &1.06 &22373& Standing up from a chair\\
			8&Standing &5587 (93) &15.56 &328655&Static standing on a solid surface\\
			9&Bicycling &2661 (44) &7.41 &156560 & Typical bicycling\\
			10&Up by elevator &1515 (25) &4.22 &89144&Standing in an elevator while moving up\\
			11&Down by elevator &1185 (19) &3.30 &69729&Standing in an elevator while moving down\\
			12&Sitting in car &3069 (51) &8.55 &180573 &Sitting while an travelling by car as a passenger\\
			\hline
			&Total &35903 &598 &100.00 &2111962 \\
			\hline
		\end{tabular}
	\end{adjustbox}
\end{table}

\section{Data Collection and Sensor Network Topology}
\label{sec:collectio&topology}
In data collection, we used MPU9250 inertial sensors and electromyography (EMG) sensors.
Each EMG sensor has a voltage gain is about 5000 and band-pass filter with bandwidth corresponding to power spectrum of EMG (10-500 Hz). A sample rate of each EMG-channel is 1.0 kHz, ADC resolution is 8 bits, input voltages: 0 - 5 Volts.
The inertial sensors consisted of a 3-axis accelerometer and a 3-axis gyroscope integrated into a single chip. Data were collected with accelerometer's range equal to $\pm 2$g with sensitivity 16.384 LSB/g and gyroscope's range equal to $\pm 2000 \degree/$s with sensitivity 16.4 LSB $/ \degree/$s. 
All sensors are powered from a battery, that helps to minimize electrical grid noise.

Accelerometer and gyroscope signals were stored in int16 format. EMG signals are stored in uint8. Therefore, accelerometer data can be converted to m~/~s$^2$ by dividing raw data 32768 and multiplying it by 2g. Raw gyroscope data can be converted to  $ \degree/$s by multiplying it by 2000/32768. 
Raw EMG data can be converted to Volts by multiplying it 0.001/255. We kept the raw data in our data collection in case one prefers other normalization techniques.

\begin{figure}[!ht] 
	\centering
	\includegraphics[width=0.5\textwidth]{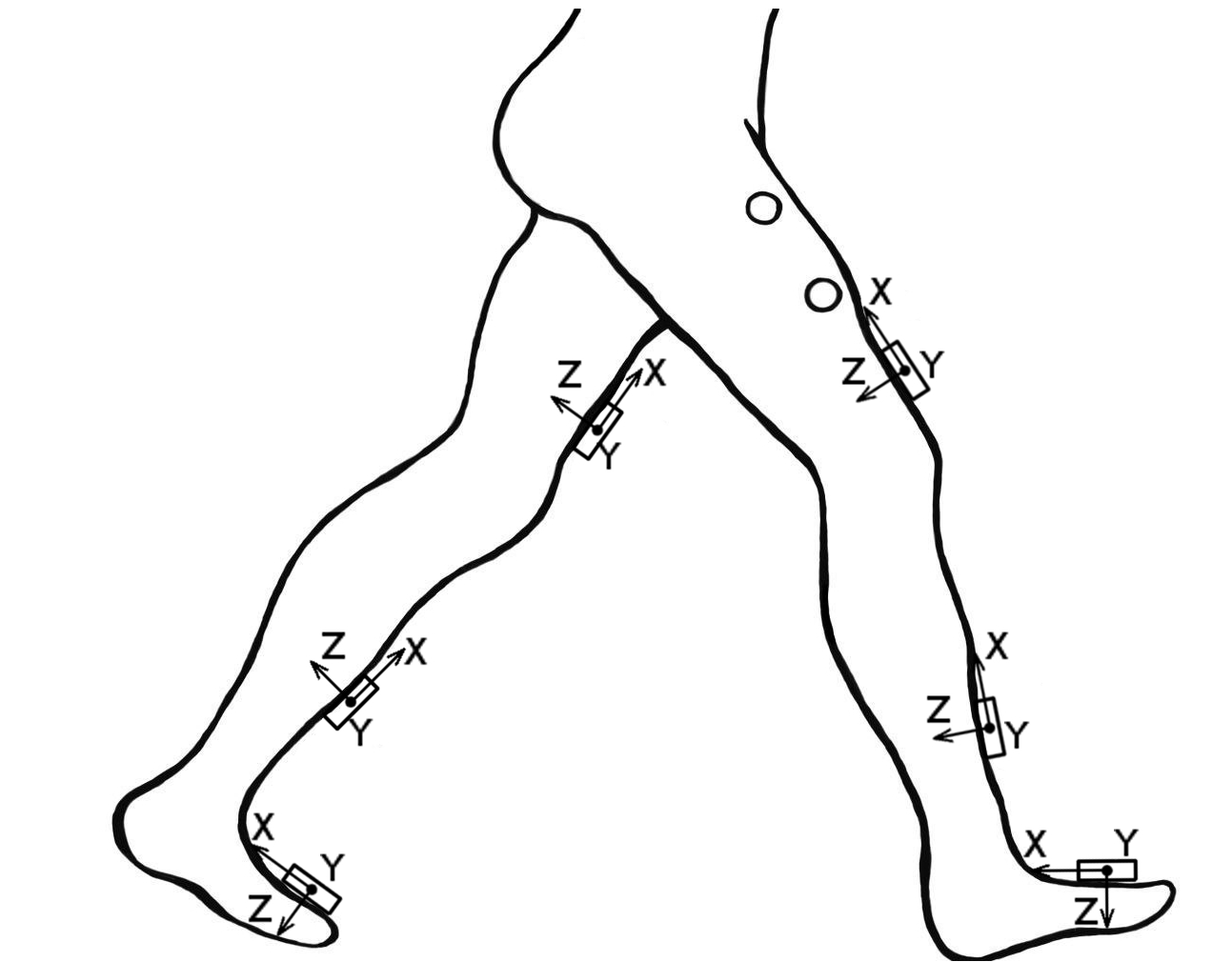}
	\caption{Location of Sensors. EMG sensor are shown as circles while boxes represent inertial sensors} 
	\label{fg:walking-scheme}
\end{figure}

In total, three pairs of inertial sensors and one pair of EMG sensors were installed symmetrically on the right and left legs with elastic bands. A pair of inertial sensors were installed on the rectus femoris muscle 5 centimetres above the knee, a pair of sensors around the middle of the shinbone at the level where the calf ends, and a pair on the feet on the metatarsal bones. Two EMG sensors were placed on vastus lateralis and connected to the skin with three electrodes. The locations of the sensors are shown in Figure \ref{fg:walking-scheme}. In total, 38 signals were collected, 36 from the inertial sensors and 2 from the EMG sensors.

The sensors were connected through wires with each other and to a microcontroller box, which contained an Arduino electronics platform with a Bluetooth module. The microcontroller collected 56.3500 samples per second in average with standard deviation (std) 3.2057 and then transmitted them to a laptop through Bluetooth connection.

The data were collected from 18 participants. These participants were healthy young adults: 4 females and 14 males, average age of 23.67 (std: 3.69) years, an average height of 179.06 (std: 9.85) cm, and an average weight of   73.44 (std: 16.67) kg.  

The participants performed a combination of activities at normal speed and casual way, and there were no obstacles placed on their way.  For instance, a participant was instructed to perform the following activities: starting from a sitting position, sitting - standing up - walking - going up the stairs - walking - sitting down. The experimenter recorded the data continually using a laptop and annotated the data with the activities performed. This provided us a long, continuous sequence of segmented data annotated with activities. We developed our own data collector program. In total, 2,111,962 samples were collected from all the 18 participants, and they provided a total of 10 hours of data.

Data acquisition was carried out mainly inside a building. However, activities such as running, bicycling, and sitting in a car were performed outside. We collected data in a moving elevator and vehicle. In these scenarios, the activities performed were simply standing or sitting. However, a force impact on the accelerometer sensors and in certain applications, it may be important to consider these facts. Note that we did not collect data on a treadmill. 

\section{Data Format}
\label{sec:format}
Data obtained from the sensors were stored in flat text files. We decided to store the data in flat files because they have one of the most universal formats, and they can be easily preprocessed in all programming languages on every system. One data file contains one recording, which is either a single activity (e.g., walking) or a series of activities. Every file name was created according to the template {\bf HGD\textunderscore vX\textunderscore ACT\textunderscore PR\textunderscore CNT.txt}. HGD is a prefix that means human gait data and vX means the version of the data files, currently v1. ACT is a variable, and it denotes the activity ID that was performed. If a file contains a series of different types of activities, then it is indicated as VARIOUS. PR indicates the ID of the person who performed the activity. Data recording was repeated a few times, and CNT is a counter for this. For example, a file named HGD\textunderscore v1\textunderscore walking\textunderscore17\textunderscore02.txt contains data from participant 17 while he was walking for the second time. The file naming convention is summarized in Table \ref{tb:filenaming}.
\begin{table}[ht!]
	\begin{center}
		\caption{Description of the file naming convention}
		\label{tb:filenaming}
		\begin{tabular}{llll}
			\hline
			TAG & Description & Type & Comment \\ 
			\hline
			HGD & Prefix &fixed & Data files start with this prefix \\
			vX & Version number &integer & Indicates the version of the data format\\
			ACT & Activity &string & Indicates the type of activity\\
			PR & Participant ID &integer & Indicates the subject whos data was recorded\\
			CNT & Counter &integer & Counter for repeated experiments\\			
			\hline
		\end{tabular}
	\end{center}
\end{table}

\begin{table}[ht!]
	\caption{Description of the data file header}
	\label{tb:metadata}
	\begin{adjustbox}{max width=\textwidth}
		\begin{tabular}{llll}
			\hline
			TAG & Description & Type & Comment \\ 
			\hline
			\#Activity & List of the activities &string & Lists the activity types in this file\\			
			\#ActivityID & List of the ID of activities &List of integers & lists the activity types in this file\\			
			\#Date-Time & Date and Time & YEAR-MM-DD-HR-MN & Year-Month-Day-Hour-Min format\\			
			\hline
		\end{tabular}
	\end{adjustbox}
\end{table}

The main body of the data files contains tab-delimited raw, unnormalized data obtained from the sensors directly. Each data file starts with a header, which contains metainformation. It summarizes the list of activities, the IDs of the activities recorded, and the time and date of the recording. This is summarized in Table \ref{tb:metadata}.

The main data body of every file has 39 columns. Each column corresponds to a sensor, and one row corresponds to a sample. The order of the columns is fixed. The first 36 columns correspond to the inertial sensors, the next 2 columns correspond to the EMG sensors, and the last column contains the activity ID. The activities are coded as shown in Table \ref{tb:activity-data-dist}. The inertial sensors are listed in the following order:  right foot (RF), right shin (RS), right thigh (RT), left foot (LT), left shin (LS), and left thigh (LT), followed by right EMG (R) and left EMG (L). Each inertial sensor produces three acceleration data on x,y,z axes and three gyroscope data on x,y,z axes. For instance, the column named 'RT\textunderscore acc\textunderscore z' contains data obtained from the z-axis of accelerometer located on the right thigh. 

\begin{figure}[!ht] 
	\centering
	\includegraphics[width=\textwidth]{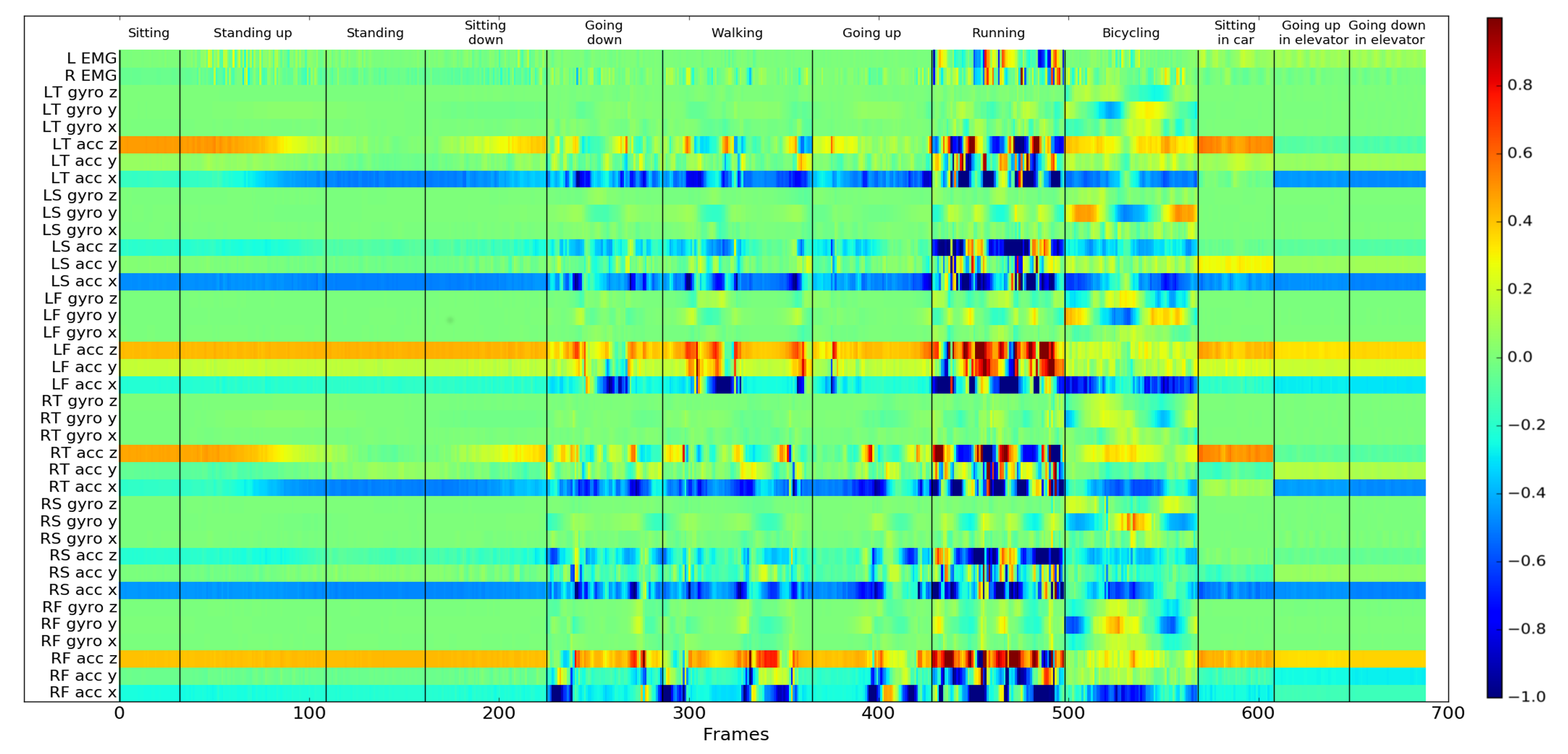}
	\caption{Data visualization. For normalization data from initial sensors were divided by 32768 and data from EMG were subtracted by 128 and divided by 128}
	\label{fg:hgd-data}
\end{figure}

Sample data with respect to the activities are visualized through a heat map representation in Figure \ref{fg:hgd-data}.

\begin{figure}[!ht] 
	\centering
	\includegraphics[width=\textwidth]{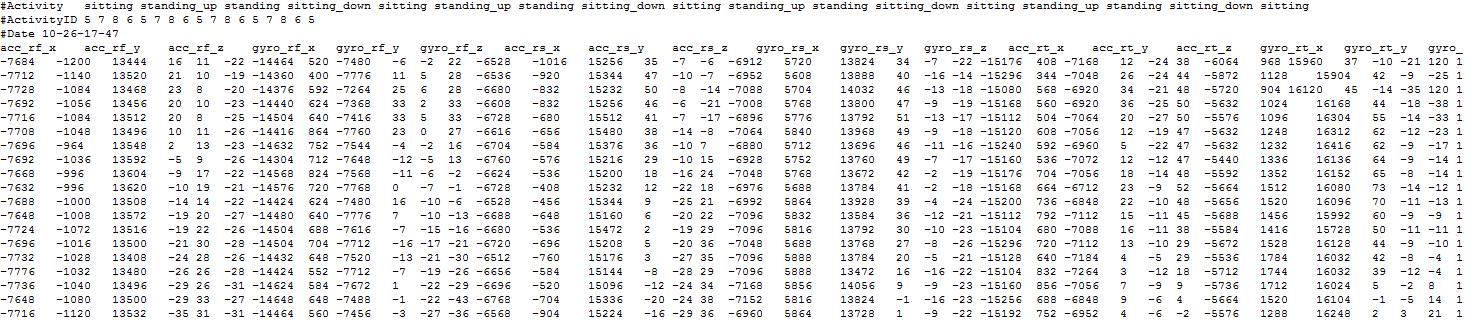}
	\caption{Screenshot of the data file}
	\label{fg:hgd-example}
\end{figure}

A screenshot of some part of data file can be seen in Figure \ref{fg:hgd-example}

The data files can be loaded easily in most of the popular programming languages. For instance, they can be loaded in Python using the following script:
\begin{lstlisting}[language=Python]
	import numpy as np 	
	data = np.genfromtxt(path_to_file, 
	                     delimiter='\t',
                             skip_header=4)
\end{lstlisting}
Please note that it requires NumPy library. It also can be loaded in Matlab with the following one-line command:

\begin{lstlisting}[language=Matlab]
	data = dlmread(path_to_file,'\t', 4,0);
\end{lstlisting}

\noindent We have prepared a script to load the data into SQLite database, which is available at the database's website:
\url{https://github.com/romanchereshnev/HuGaDB/blob/master/Scripts/create_db.py}.

\section{Discussion on Data Variance}

\begin{figure}[tbp]
	\centering
	\small
	\resizebox{\textwidth}{!}{
		\begin{tabular}{cc}
			\includegraphics[width=\textwidth]{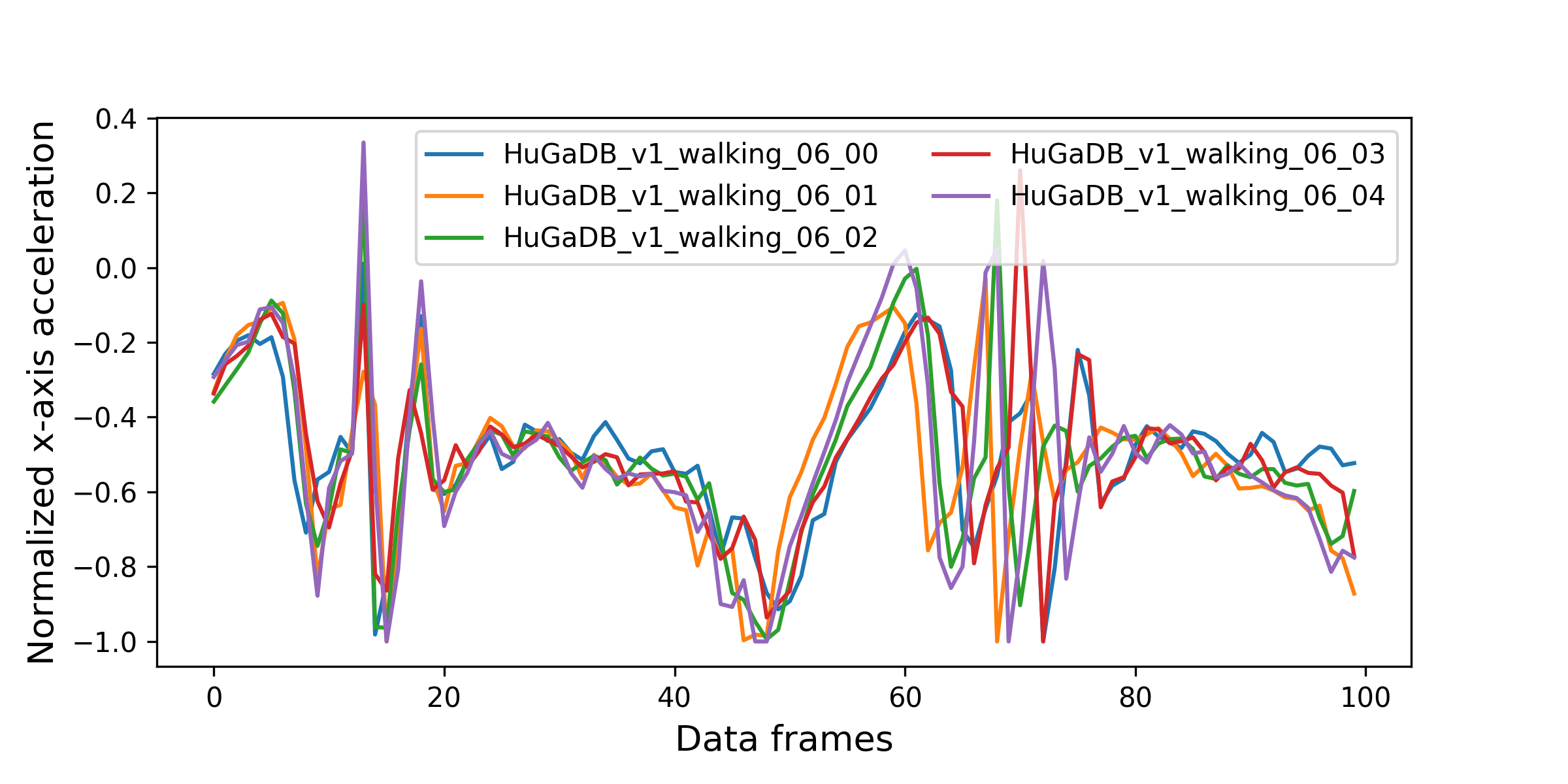}\\
			(A) Single user\\
			\includegraphics[clip=true,width=\textwidth]{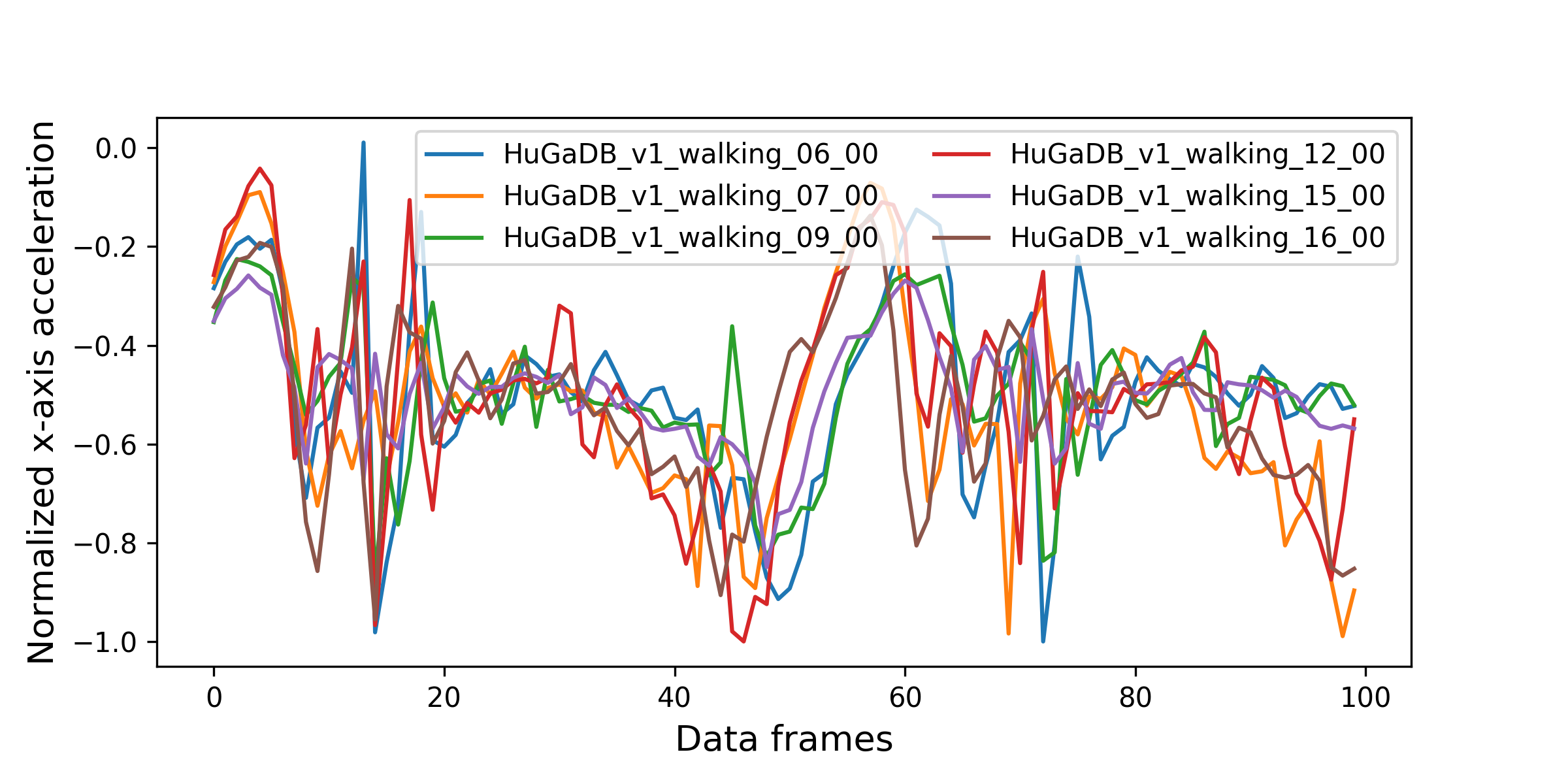}\\
			(B) Various users\\
		\end{tabular}	
	}
	\caption{Data variance during walking. A) Activity performed by the same user multiple times. B) Activity performed by different users. Legend indicates the source of the data. Data are scaled to the range $[-1,+1]$.\label{fig:variance}}
\end{figure}

We were interested seeing the variance among the data collected, in particular, the data variance A) within a single user and B) between several users. For this reason, we plotted in Figure \ref{fig:variance} the x-axes acceleration data from the thigh recorded during a short two-three-step walk. Panel A shows the data from various recordings performed by the same user. It can be seen that the data variance at a single frame is quite low suggesting that people perform activities very similar way. On the other hand, panel B shows data obtained from six different, randomly chosen users. Here, a much higher variance can be seen in the same frames compared to the previous case. The increased variance may arise from several facts including: difference in gait, difference in leg shape, sensors mounted in slightly different positions, etc. We obtained similar conclusions on data obtained from different sensors during different activities. We note that, even higher variance was observed in the EMG data, which resulted from the difference in the electricity conduction characteristics of the skin, skin thickness, etc.

Taking into account the high data variance between different users, we emphasize the importance of proper evaluation of machine learning methods developed for human activity recognition. Therefore, we propose using the supervised cross-validation approach for constructing training and test sets \cite{kertesz2008benchmarking}. In this approach, all the data from a designated user are held out only for tests and the data from the other 17 participants are used for training. Thus, this approach provides a reliable estimation of how an activity recognition system would perform with a new user whose data was not seen before.

Variance can arise from using different brands of sensors. Unfortunately, we did not have the capacity to collect data from different brands of sensors. We hope the measurement noise is small in general and that different sensors can be calibrated to be compatible with each other.



\section{Availability}
\label{sec:availability}
The database is available free of charge at \url{https://github.com/romanchereshnev/HuGaDB} (455 Mb).

\section{Summary}
\label{sec:summery}
The HuGaDB dataset contains detailed kinematic data for analyzing human gait and activity recognition. This dataset differs from previously published datasets in the sense that HuGaDB provides human gait data in great detail mainly from inertial sensors and contains segmented annotations for studying the transition between different activities. Data were obtained from 18 participants, and in total, they provide around 10 hours of recording. This dataset can be used in health-care-related studies, such as walking rehabilitation, or in modeling human movements in virtual reality or humanoid robotics. The dataset will be updated with new data from new participants in the future.



\bibliographystyle{unsrt}
\bibliography{bibliography}

\end{document}